\documentclass[aps,prd,twocolumn,superscriptaddress]{revtex4}

\begin{document}


\title{Does a black hole rotate  
       in Chern-Simons modified gravity?}


\author{Kohkichi Konno}
\email[]{konno@topology.coe.hokudai.ac.jp}
\affiliation{Department of Applied Physics, Hokkaido University,
Sapporo 060-8628, Japan}
\author{Toyoki Matsuyama}
\affiliation{Department of Physics, Nara University of Education,
Nara 630-8528, Japan}
\author{Satoshi Tanda}
\affiliation{Department of Applied Physics, Hokkaido University,
Sapporo 060-8628, Japan}


\date{\today}

\begin{abstract}
Rotating black hole solutions in the (3+1)-dimensional 
Chern-Simons modified gravity theory are discussed 
by taking account of perturbation around the Schwarzschild solution.
The zenith-angle dependence of a metric function
related to the frame-dragging effect is determined
from a constraint equation independently of a choice 
of the embedding coordinate.
We find that at least within the framework of 
the first-order perturbation method,
the black hole cannot rotate for finite black hole mass
if the embedding coordinate is taken to be a timelike vector. 
However, the rotation can be permitted
in the limit of $M/r \rightarrow 0$ 
(where $M$ is the black hole mass and $r$ is the radius).
For a spacelike vector, the rotation can also be permitted
for any value of the black hole mass.
\end{abstract}


\maketitle

\section{Introduction}

The latest observational results of the cosmic microwave 
background (CMB) anisotropy from the Wilkinson Microwave Anisotropy 
Probe (WMAP) \cite{wmap} are successfully explained by the 
$\Lambda$ cold dark matter standard model.
However, two big issues still remain: what is dark matter,
and what is dark energy?
According to the WMAP results, unfortunately 
about 96\% of the contents of the universe 
is given by the dark components that we still do not know.
Therefore, properties of dark matter and 
dark energy have eagerly been investigated 
from observations \cite{g-lens,type-ia}.

In contrast with the ordinary approaches \cite{g-lens,type-ia}, 
in which the existence of the dark components is assumed, 
it is of great interest to investigate
alternative gravity theories \cite{mond,beken,cdtt,cct,no-o}
to solve the dark-matter and dark-energy problem.
In this paper, we focus our attention on 
the Chern-Simons modified gravity theory \cite{djt}.
This gravity theory was constructed by Deser et al.~\cite{djt} 
in (2+1) spacetime dimensions for the first time 
by analogy with the topologically massive U(1) and SU(2) 
gauge theories.
The Chern-Simons modified gravity theory
was relatively recently extended by Jackiw \& Pi \cite{jp}
to (3+1) spacetime dimensions. 
In the extended theory, the Schwarzschild solution 
holds without any modification \cite{jp}.
Therefore, the theory passes the classical tests
of general relativity \cite{weinberg}.
In this gravity theory, however,  
the Kerr solution does not hold.
Thus, the solution for a rotating black hole should have 
a different form from the Kerr solution. 
In (2+1)-dimensional Chern-Simons modified gravity, a family of rotating 
black hole solutions was found by Moussa et al.~\cite{mcl}.
The solutions have a fascinating feature that 
observers in this spacetime behave like ones inside 
the ergosphere of the Kerr spacetime.
This feature is similar to that of the rotation of 
galaxies \cite{g-rotation}. Therefore, it is very interesting to 
investigate rotating black hole solutions in 
the (3+1)-dimensional Chern-Simons modified gravity theory.
In this paper, we discuss rotating black hole solutions
taking account of perturbation around the Schwarzschild solution.

This paper is organized as follows.
In Sec.~\ref{sec:cs-review}, we briefly review
the (3+1)-dimensional Chern-Simons modified gravity theory.
In Sec.~\ref{sec:pertub}, we consider the perturbation
around the Schwarzschild solution to discuss slow rotation
of the black hole. First we investigate a constraint equation
independently of a choice of the embedding coordinate. 
In Sec.~\ref{sec:pertub-t}, from the first-order equations 
of the field equation, we obtain the metric solution
taking the embedding coordinate to be timelike.
In Sec.~\ref{sec:pertub-s}, we investigate the metric solution 
for the case in which the embedding coordinate is spacelike.
Finally, we provide a summary in Sec.~\ref{sec:summary}.
In this paper, we use a unit in which $c=G=1$.

\section{Brief Review of Chern-Simons Modified Gravity Theory}
\label{sec:cs-review}

We briefly review the Chern-Simons modification of 
general relativity developed by Jackiw \& Pi \cite{jp}.
The Chern-Simons modified gravity theory is provided by
the action
\begin{eqnarray}
 I & = & \int dx^4 {\cal L} 
 \: = \: \frac{1}{16\pi} \int dx^4 
  \left( \sqrt{-g} R + \frac{1}{4} 
   \vartheta \: ^{\ast} \!R R \right) \nonumber \\
 & = &\! \frac{1}{16\pi} \int dx^4 
  \left( \sqrt{-g} R - \frac{1}{2} 
   v_{\mu} K^{\mu} \right) , 
\end{eqnarray}
where the first term in the integrand is the Einstein-Hilbert
action, and $v_{\mu} \equiv \partial_{\mu} \vartheta$ is an external
4-vector, which is called the embedding coordinate. 
The Pontryagin density $^{\ast} \!R R$ is defined by
$^{\ast} \!R R \equiv \; \! \!^{\ast} \!R^{\sigma\ \mu\nu}_{\ \tau} 
R^{\tau}_{\ \sigma\mu\nu}$, using the dual Riemann tensor 
$^{\ast} \!R^{\tau\ \mu\nu}_{\ \sigma} \equiv 
\frac{1}{2} \varepsilon^{\mu\nu\alpha\beta} 
R^{\tau}_{\ \sigma\alpha\beta}$.
The Chern-Simons topological current $K^{\mu}$ is given by 
\begin{equation}
 K^{\mu} = \varepsilon^{\mu\alpha\beta\gamma}
  \left[ \Gamma^{\sigma}_{\alpha\tau}
  \partial_{\beta} \Gamma^{\tau}_{\gamma\sigma} 
  + \frac{2}{3} \Gamma^{\sigma}_{\alpha\tau}
  \Gamma^{\tau}_{\beta\eta} \Gamma^{\eta}_{\gamma\sigma} \right] ,
\end{equation}
which is related to the Pontryagin density as
$\partial_{\mu} K^{\mu} = \frac{1}{2} \; \! ^{\ast} \!R R$.

 From the variation of the Lagrange density ${\cal L}$
with respect to the metric $g_{\mu\nu}$, 
it turns out that the field equation has the form
\begin{equation}
 \label{eq:feq}
 G^{\mu\nu} + C^{\mu\nu} = -8\pi T^{\mu\nu} ,
\end{equation}
where $G^{\mu\nu} \equiv R^{\mu\nu} - \frac{1}{2} g^{\mu\nu} R$
is the Einstein tensor, $T^{\mu\nu}$ is the energy-momentum tensor,
and $C^{\mu\nu}$ is the Cotton tensor defined as
\begin{eqnarray}
 C^{\mu\nu} & = & - \frac{1}{2\sqrt{-g}}
 \left[ v_{\sigma} \left( \varepsilon^{\sigma\mu\alpha\beta}
   \nabla_{\alpha} R^{\nu}_{\ \beta} + 
   \varepsilon^{\sigma\nu\alpha\beta}
   \nabla_{\alpha} R^{\mu}_{\ \beta} \right) \right.
   \nonumber \\
 && \qquad \qquad + \left. v_{\tau \sigma}
   \left( \:\! ^{\ast} R^{\tau\mu\sigma\nu} 
    + \:\! ^{\ast} R^{\tau\nu\sigma\mu} \right)\right] .
\end{eqnarray}
Here $v_{\tau \sigma} \equiv \nabla_{\sigma} v_{\tau} 
= \partial_{\sigma} \partial_{\tau} \vartheta
 - \Gamma^{\lambda}_{\tau\sigma} \partial_{\lambda} \vartheta$
is a symmetric tensor.
Corresponding to the Bianchi identity $\nabla_{\mu} G^{\mu\nu} = 0$
and the equation of motion $\nabla_{\mu} T^{\mu\nu} = 0$,
the following condition should be imposed:
\begin{equation}
\label{eq:cndt}
 0 = \nabla_{\mu} C^{\mu\nu} 
  = \frac{1}{8\sqrt{-g}} v^{\nu} \;\! ^{\ast} R R .
\end{equation}
This constraint equation implies that diffeomorphism symmetry 
breaking is suppressed \cite{jp}.

\section{Perturbative Approach to Rotating Black Hole Solutions}
\label{sec:pertub}

In the Chern-Simons modified gravity theory, 
the Schwarzschild solution as a non-rotating black hole solution 
holds without any modification as mentioned above.
The Schwarzschild metric is given by
\begin{eqnarray}
\label{eq:sws}
 ds^2 & = & - \left( 1- \frac{2M}{r} \right) dt^2
  + \left( 1- \frac{2M}{r} \right)^{-1} dr^2 \nonumber \\
 &&  + r^2 \left( d\theta^2 + \sin^2 \theta d\phi^2 \right) , 
\end{eqnarray}
where $M$ is the black hole mass.
This solution gives $C^{\mu\nu} = 0$ and
$\nabla_{\mu} C^{\mu\nu} = v^{\nu} \:\! ^{\ast}RR 
/\left( 8\sqrt{-g} \right) = 0$ 
trivially.

In order to take account of rotation of the black hole, 
let us consider perturbation around the Schwarzschild solution.
In the perturbation, the expansion parameter $\epsilon$ is
related to the angular momentum $J$ of the black hole,
i.e., $J \sim O\left( \epsilon\right)$.
Under the assumption of stationary, axisymmetric spacetime, 
we can write the form of the perturbed metric 
as \cite{rw,hs,hartle}
\begin{eqnarray}
\label{eq:psws}
 ds^2 & = & - \left( 1- \frac{2M}{r} \right) 
  \left( 1 + h\left(r,\theta \right)\right) dt^2 \nonumber \\
 &&  + \left( 1- \frac{2M}{r} \right)^{-1} 
  \left( 1 + m\left(r,\theta \right)\right) dr^2 \nonumber \\
 &&  + r^2 \left( 1 + k\left(r,\theta \right)\right) \nonumber \\
 && \quad  \times
  \left[ d\theta^2 + \sin^2 \theta 
  \left( d\phi - \omega \left( r,\theta \right) dt \right)^2 \right] . 
\end{eqnarray}
The functions $h\left( r,\theta\right)$, $m\left( r,\theta\right)$, 
$k\left( r,\theta\right)$ and $\omega\left( r,\theta\right)$
are of the first order in $\epsilon$.
Hereafter, we take account of equations 
up to the first order in $\epsilon$.

Using this perturbed metric, from the condition (\ref{eq:cndt}),
we obtain 
\begin{eqnarray}
 0 & = & \nabla_{\mu} C^{\mu\nu} \nonumber \\
 & = & v^{\nu} \frac{3M}{r^3} \sin \theta \;
     \left[ \omega_{, r\theta} + 2\cot\theta \: \omega_{, r} \right] ,
\end{eqnarray}
where a subscript comma denotes the partial differentiation
with respect to the coordinates.
In this expression, 
the function $\omega \left( r,\theta \right)$
only appears. Therefore, we find that solutions for 
$\omega \left( r,\theta \right)$ should have 
the functional form
\begin{equation}
\label{eq:omega}
 \omega \left( r , \theta \right) = \frac{\varpi (r)}{\sin^2 \theta} ,
\end{equation}
where $\varpi$ is a function of $r$ only.
While $\omega \left( r , \theta \right)$ 
is singular on the rotation axis ($\theta = 0$ and $\pi$), 
the metric is regular at least up to the first order, because
$g_{t\phi} = - r^2 \varpi (r)+ O(\epsilon^2 )$.
Note that $g_{t\phi}$ does not vanish on the rotation axis
unless $\varpi (r)$ is identically zero.
This means that the shift vector 
$N_{i} \equiv g_{ti} \: (i=r, \theta , \phi )$
defined in the (3+1) formalism \cite{adm}
is singular on the rotation axis.
It should also be noted that this result is independent 
of a choice of the embedding coordinate $v^{\nu}$.

\subsection{Linear perturbation equations 
            and the metric solution for timelike $v_{\mu}$}
\label{sec:pertub-t}

We adopt a timelike vector for $v_{\mu}$, i.e.,
$v_{\mu} = \left( 1/\mu_{0} , 0,0,0 \right)$,
which is derived from $\vartheta = t / \mu_{0}$.
In our universe, there exists the frame of reference
in which the CMB radiation
can be seen as an isothermal distribution except
for small fluctuations.
The frame of reference is specified by a timelike
vector. Such a timelike vector is a candidate
for the timelike vector $v_{\mu}$.

 From the ($tt$), ($rr$), ($r\theta$), ($r\phi$), 
($\theta\theta$), ($\theta\phi$), and ($\phi\phi$)-components 
of Eq.~(\ref{eq:feq}), 
we can obtain homogeneous differential equations
for the functions $h$, $m$ and $k$. Hence, 
the homogeneous differential equations have a simple solution 
of $h( r,\theta ) = m( r,\theta ) = k( r,\theta ) =0$.
These differential equations are completely decoupled from 
the function $\omega$. 
Since we are now interested in the rotation of the black hole, 
i.e., the function $\omega$, we do not seek any other solutions 
for the functions $h$, $m$ and $k$.
 From the ($tr$), ($t\theta$) and 
($t\phi$)-components of Eq.~(\ref{eq:feq}), we 
obtain the equations for $\omega$
\begin{eqnarray}
\label{eq:feq1}
 0 & = & \omega_{, \theta\theta\theta} 
 + r(r-2M) \omega_{, rr\theta} 
 + 2r(r-2M) \cot\theta \omega_{,rr} \nonumber \\
 && + \: 5 \cot \theta \omega_{,\theta\theta}
 + 2(2r-5M) \omega_{,r\theta} \nonumber \\
 && + \: 4(2r-5M) \cot\theta \omega_{,r} 
 + 3 \left( \cot^2 \theta -1 \right) \omega_{,\theta} ,
\end{eqnarray}
\begin{eqnarray}
\label{eq:feq2}
 0 & = & r^2 (r-2M) \omega_{, rrr} + r \omega_{,r\theta\theta} 
 + 3r\cot\theta \omega_{,r\theta} \nonumber \\
 && + \: 6r (r-2M) \omega_{,rr} + 4(r-3M) \omega_{,r} ,
\end{eqnarray}
\begin{eqnarray}
\label{eq:feq3}
 0 & = & r (r-2M) \omega_{, rr} + 4(r-2M) \omega_{,r}
  + \omega_{,\theta\theta} \nonumber \\ 
 && + \: 3\cot\theta \omega_{,\theta} .
\end{eqnarray}
Here, Eqs.~(\ref{eq:feq1}) and (\ref{eq:feq2}) are
obtained, respectively, only from the non-vanishing ($tr$) and 
($t\theta$)-components of the Cotton tensor,
and Eq.~(\ref{eq:feq3}) is obtained only from the 
($t\phi$)-component of the Einstein tensor.
Note that these equations do not include $\mu_{0}$.
This is due to a shortcoming of 
the first-order perturbation method.

Using the result of Eq.~(\ref{eq:omega}), 
Eq.~(\ref{eq:feq1}) is automatically satisfied.
 From Eqs.~(\ref{eq:feq2}) and (\ref{eq:feq3}),
we obtain the differential equations
for $\varpi (r)$, respectively, 
\begin{equation}
\label{eq:deq-o1}
 r^2 \varpi''' + 6r \varpi'' + 6\varpi' = 0 ,
\end{equation}
\begin{equation}
\label{eq:deq-o2}
 r(r-2M) \varpi'' + 4(r-2M) \varpi' + 2\varpi = 0 ,
\end{equation}
where a prime denotes the differentiation with respect 
to the coordinate $r$.
The solution of Eq.~(\ref{eq:deq-o1}) is given by
\begin{equation}
\label{eq:solution1}
 \varpi = C_{0} + \frac{C_{1}}{r} + \frac{C_{2}}{r^2} ,
\end{equation}
where $C_{0}$, $C_{1}$ and $C_{2}$ are constants of integration.
On the other hand, the solution of Eq.~(\ref{eq:deq-o2}) 
is given by
\begin{eqnarray}
\label{eq:solution2}
 \varpi & = & D_{1} \frac{r-2M}{r^3} 
  + \frac{D_{2} }{r^3} \left[ r^2 - 2Mr -4M^2 \right.
   \nonumber \\ 
 &&  \qquad \qquad 
  + \left. 4M \left( r-2M \right) \ln (r-2M) \right] ,
\end{eqnarray}
where $D_{1}$ and $D_{2}$ are constants of integration.
Thus, the solution that satisfies both 
differential equations (\ref{eq:deq-o1}) and (\ref{eq:deq-o2})
is given only by $\varpi (r) = 0$.
Therefore, we conclude that 
within the framework of the first-order perturbation,
the black hole cannot rotate in the Chern-Simons 
modified gravity for the timelike vector.

However, in the limit of $M/r \rightarrow 0$, 
the derivative of Eq.~(\ref{eq:deq-o2}) coincides with 
Eq.~(\ref{eq:deq-o1}). Hence, the solution
\begin{equation}
\label{eq:l-s-w}
 \varpi = \frac{C_{1}}{r} + \frac{C_{2}}{r^2} ,
\end{equation}
i.e.,
\begin{equation}
\label{eq:l-s-g}
 g_{t\phi} = - \left( C_{1} r + C_{2} \right),
\end{equation}
is permitted in this limit. 
Since the metric component $g_{t\phi}$ is proportional to $r$
at infinity, the frame-dragging effect of this solution
works in the whole space.

\subsection{Linear perturbation equations and the
            metric solution for spacelike $v_{\mu}$}
\label{sec:pertub-s}

Next we take another choice of $\vartheta = r\cos\theta / \lambda_{0}$.
This provides a spacelike vector 
$v_{\mu} = \left( 0 , \cos\theta / \lambda_{0} , 
 - r \sin\theta / \lambda_{0} , 0 \right)$,
which becomes a unit vector parallel to the rotation axis at infinity. 
The discrepancy between the observational result and the theoretical 
prediction in the quadrupole moment of the CMB anisotropy
may imply the existence of such a spacelike vector \cite{cct-cmb}.

 From the $(tt)$, $(t\phi)$, $(rr)$, $(r\theta)$
$(\theta\theta )$, and $(\phi\phi )$-components 
of the field equation, we obtain the non-vanishing
equations, respectively, 
\begin{eqnarray}
\label{eq:sp-tt}
 \lefteqn{2r(r-2M) k_{,rr} - 2(r-2M) m_{,r} + 2(3r-5M) k_{,r}}
         \nonumber \\
 \lefteqn{\ + \: k_{,\theta\theta} + \cot\theta k_{,\theta}
    + \: m_{,\theta\theta} + \cot\theta m_{,\theta}
    + 2k-2m} \nonumber \\
 & = & - \: \frac{1}{\lambda_{0} (r-2M)}
  \left[ r^2 (r-2M)^2 \varpi''' \right.  \nonumber \\
 && \qquad + \: r(r-2M)(6r-11M) \varpi'' 
  \nonumber \\
 && \qquad \left. 
  + \: 2 (r-2M)(3r-2M) \varpi' - 2M \varpi \right] , \\
\label{eq:sp-tp}
 \lefteqn{\frac{1}{\sin^2 \theta}
  \left[ r(r-2M) \varpi'' + 4(r-2M) \varpi' + 2\varpi \right]}
  \nonumber \\
 &= & - \: \frac{1}{2 \lambda_{0} r^3} \left[ 
  -r(r-2M)^2 \left( r h_{,rrr} + \cot\theta \: h_{,rr\theta} \right)
  \right. \nonumber \\
 && + r(r-2M)^2 \left( r k_{,rrr} 
    + \cot\theta \: k_{,rr\theta} \right) 
    \nonumber \\
 && - r(r-2M) \left( h_{,r\theta\theta} - k_{,r\theta\theta}\right)
    \nonumber \\
 && -(r-2M) \cot\theta \left( h_{,\theta\theta\theta}
    - k_{,\theta\theta\theta} \right)
    -r(r-2M) \nonumber \\
 && \ \times \left\{ (2r+M) h_{,rr} + (r-3M) m_{,rr}  
    - (3r-2M) k_{,rr} \right\}
    \nonumber \\
 && - (r-2M) \cot\theta \nonumber \\
 && \ \times \left\{ 
    (3r+M) h_{,r\theta} - (3r-2M) k_{,r\theta} 
    - 3M m_{,r\theta} \right\}
    \nonumber \\
 && + (2r-5M) h_{,\theta\theta}
    -(r-3M) m_{,\theta\theta} -(r-2M) k_{,\theta\theta}\nonumber \\
 && -(r-2M) \cot^2 \theta \left( h_{,\theta\theta} 
    - k_{,\theta\theta} \right)\nonumber \\
 && + \left( 2r^2 - 9Mr + 10M^2 \right) h_{,r}
    -9M(r-2M)m_{,r}  \nonumber \\
 && -2 \left( r^2 - 9Mr + 14M^2 \right) k_{,r}
    \nonumber \\
 && +  3(r-3M) \cot\theta 
  \left( h_{,\theta} - m_{,\theta} \right)
  \nonumber \\
 && \left. + (r-2M) \cot^3\theta 
  \left( h_{,\theta} - k_{,\theta} \right) 
  + 2(r-2M) (m-k) \right] , \nonumber \\
&& \\
\label{eq:sp-rr}
\lefteqn{(r-2M) \left[ h_{,\theta\theta}
  + \cot\theta h_{,\theta} + k_{,\theta\theta}
  + \cot\theta k_{,\theta}\right.} \nonumber \\
\lefteqn{\qquad \qquad \quad \left. 
 + 2(r-2M) (h_{,r}+k_{,r}) +2(k-m) \right]} \nonumber \\
& = & \frac{r-3M}{\lambda_{0}}
  \left[ r(r-2M) \varpi'' + 4(r-2M) \varpi' + 2\varpi \right] , 
  \\
\label{eq:sp-rth}
\lefteqn{r(r-2M) \left( h_{,r\theta} + k_{,r\theta} \right)
  -(r-3M)h_{,\theta}-(r-M)m_{,\theta} } \nonumber \\
& = &  - \frac{r\cot\theta}{\lambda_{0}}
  \left[ r(r-2M) \varpi'' + 4(r-2M) \varpi' + 2\varpi \right] , \\
\label{eq:sp-thth}
\lefteqn{r(r-2M) \left( h_{,rr} + k_{,rr} \right)
  +(r+M)h_{,r}+2(r-M)k_{,r} } \nonumber \\
\lefteqn{\quad - (r-M) m_{,r} + \cot\theta 
    \left( h_{,\theta} + k_{,\theta} \right)} \nonumber \\
& = & \frac{\cot^2 \theta}{\lambda_{0}}
  \left[ r(r-2M) \varpi'' + 4(r-2M) \varpi' + 2\varpi \right] , \\
\label{eq:sp-phph}
\lefteqn{r(r-2M) \left( h_{,rr} + k_{,rr} \right)
  + h_{,\theta\theta} + m_{,\theta\theta} }\nonumber \\
\lefteqn{\quad +(r+M)h_{,r} + 2(r-M)k_{,r} - (r-M) m_{,r} } \nonumber \\
& = & - \frac{1}{\lambda_{0}\sin^2\theta}
  \left[ r(r-2M) \varpi'' + 4(r-2M) \varpi' + 2\varpi \right. 
  \nonumber \\
&& \left. - r \left\{ r(r-2M) \varpi''' + 2(3r-5M)  \varpi'' 
      + 6\varpi' \right\} \sin^2 \theta \right] . \nonumber \\
\end{eqnarray}
While the right-hand side of Eq.~(\ref{eq:sp-rth}) has 
the zenith-angle dependence of $\cot\theta$, the left-hand
side is composed of the first-order derivatives 
of $h$, $m$ and $k$ with respect to $\theta$.
Thus the solution has the form of 
$(h, m, k) \propto \ln (\sin\theta )$.
However, these functions become singular along the 
rotation axis. Hence, the zenith-angle dependence 
of the functions $h$, $m$ and $k$ should vanish, 
and therefore these functions depend on $r$ only.
On the other hand, this result, i.e., 
$h=h(r)$, $m=m(r)$ and $k=k(r)$, conflicts with 
Eq.~(\ref{eq:sp-thth}), since
the left-hand side becomes a function of $r$ only, 
and the right-hand side has the dependence of $\cot^2 \theta$.
Therefore, the functions $h$, $m$ and $k$ should vanish.
Then, we derive the differential equations
\begin{eqnarray}
 \label{eq:sp-w1}
 0 & = &  r^2 (r-2M)^2 \varpi''' + r(r-2M)(6r-11M) \varpi'' 
  \nonumber \\
 &&
  + \: 2 (r-2M)(3r-2M) \varpi' - 2M \varpi , \\
\label{eq:sp-w2}
 0 & = & r(r-2M) \varpi'' + 4(r-2M) \varpi' + 2\varpi , \\
\label{eq:sp-w3}
 0 & = &  r(r-2M) \varpi''' + 2(3r-5M)  \varpi'' 
      + 6\varpi' . 
\end{eqnarray}
Equations (\ref{eq:sp-w1}) and (\ref{eq:sp-w3}) 
can be derived consistently from Eq.~(\ref{eq:sp-w2}). 
Thus the equation that we have to solve is  
Eq.~(\ref{eq:sp-w2}).
In the same way as the case for the timelike $v_{\mu}$,
the differential equation does not include the parameter
$\lambda_{0}$.
The solution of Eq.~(\ref{eq:sp-w2}) is given by
the same expression as Eq.~(\ref{eq:solution2}), 
which leads to 
\begin{eqnarray}
\label{eq:g_tp-sp}
 g_{t\phi} & = & \tilde{D}_{1} \frac{r-2M}{r} 
  + \frac{\tilde{D}_{2} }{r} \left[ r^2 - 2Mr -4M^2 \right.
   \nonumber \\ 
 &&  \qquad \qquad 
  + \left. 4M \left( r-2M \right) \ln (r-2M) \right] ,
\end{eqnarray}
where $\tilde{D}_{1}$ and $\tilde{D}_{2}$ are constants.
Therefore, for the spacelike vector $v_{\mu}$,
the spacetime rotation is permitted for any value of the black hole mass.
However, if $\tilde{D}_{2} \neq 0$, then
the frame-dragging effect extends to infinity,
because the second term in Eq.~(\ref{eq:g_tp-sp}) diverges
as $r$ increases. 
Furthermore, the result of Eq.~(\ref{eq:g_tp-sp}) means 
that the above-mentioned 
string singularity of the shift vector $N_{i}$ extends 
to infinity even if $\tilde{D}_{2}=0$.

\section{Summary}
\label{sec:summary}

We have investigated the rotation of a black hole in the Chern-Simons
modified gravity theory.
In particular, we have considered slow rotation of a 
black hole using the perturbation method, in which 
the Schwarzschild solution was taken to be the background.
 From the constraint equation, we obtained 
the zenith-angle dependence of the metric function 
$\omega (r,\theta )$ related to the frame-dragging effect,
independently of a choice of the embedding coordinate $v_{\mu}$.  
Furthermore, by solving the field equation, 
we found that the black hole cannot rotate 
for the timelike vector $v_{\mu}$ at least 
within the framework of the first-order perturbation method.
However, in the limit of $M/r \rightarrow 0$, 
the spacetime rotation is permitted, 
whose frame-dragging effect extends to infinity. 
In contrast, for the spacelike vector $v_{\mu}$,
the spacetime rotation is permitted for any value 
of the black hole mass.  
Its frame-dragging effect also extends to infinity. 
Therefore, it is still an open question
which form of the metric corresponds to the Kerr solution,
which reduces to the Minkowski metric at infinity.
Derivation of exact solutions for stationary, axisymmetric spacetimes
in the Chern-Simons modified gravity theory may solve this problem.
Then, we could also understand effects of the parameter
$\mu_{0}$ or $\lambda_{0}$, which appears 
in the Chern-Simons term, on the black hole physics.
The derivation of exact solutions will be a future work.
Furthermore, it should be noted that
the above-mentioned results might be modified by the 
extension of the theory in which $\vartheta$ in the Chern-Simons 
term is taken to be a dynamical variable.
This will also be discussed elsewhere.

\begin{acknowledgments}
This work was supported in part by a Grant-in-Aid
for Scientific Research from The 21st Century COE
Program ``Topological Science and Technology''.
Analytical calculations were performed in part
by Mathematica (Wolfram Research, Inc.)
on computers at YITP in Kyoto University.
\end{acknowledgments}


\end{document}